\documentclass[12pt]{article}
\usepackage{epsfig}
\usepackage{times}
\usepackage{rotate}

\newenvironment{Eqnarray}%
   {\arraycolsep 0.14em\begin{eqnarray}}{\end{eqnarray}}
\def\beqa{\begin{Eqnarray}}
\def\eeqa#1{\label{#1}\end{Eqnarray}}
\def\eeqan{\end{Eqnarray}}

\newcommand{\lt}{\left}
\newcommand{\rt}{\right}
\newcommand{\ov}{\overline}

\newcommand{\nn}{\nonumber \\}
\newcommand{\no}{\nonumber }
\newcommand{\bbs}{$B_s\!-\!\ov{B}{}_s\,$}
\newcommand{\bbms}{$B_s\!-\!\ov{B}{}_s\,$\ mixing}

\newcommand{\dm}{\ensuremath{\Delta m}}
\newcommand{\dg}{\ensuremath{\Delta \Gamma}}
\newcommand{\eq}[1]{(\ref{#1})}
\newcommand{\guntf}{\ensuremath{ \Gamma  [f,t] }}
\newcommand{\bra}[1]{\langle \, #1 \, | }
\newcommand{\ket}[1]{| \, #1 \, \rangle }
\newcommand{\lqcd}{\Lambda_{\textit{\scriptsize{QCD}}}}
\newcommand{\dbt}{\ensuremath{|\Delta B|  = 2}}
\newcommand{\dbo}{\ensuremath{|\Delta B|  = 1}}
\newcommand{\Bsun}{\ensuremath{\stackrel{\scriptscriptstyle \lt(
      -\rt)}{B}_s}}
\newcommand{\brunts}[1]{\ensuremath{Br \, \big(\!\Bsun \rightarrow #1 \,\big)}}
\newcommand{\epm}[2]{
 \raisebox{-0.5ex}{\shortstack[l]{$\scriptstyle+#1$\\$\scriptstyle-#2$}}}

\newcommand\pubnumber{CERN--TH/2001-135}
\newcommand\pubdate{May 22, 2001}
\newcommand\hepnumber{hep-ph/0105215}

\def\csumb{CERN -- Theory Division, 1211 Geneva 23, Switzerland}
\def\Title#1{\begin{center} {\Large\bf #1 } \end{center}}
\def\Author#1{\begin{center}{ \sc #1} \end{center}}
\def\Address#1{\begin{center}{ \it #1} \end{center}}

\newcommand\pubblock{\rightline{\begin{tabular}{l} \pubnumber\\
         \pubdate\\ \hepnumber \end{tabular}}}
\newenvironment{Abstract}{\begin{quotation}  }{\end{quotation}}
\newenvironment{Presented}{\begin{quotation} \begin{center} 
             Presented at the\end{center}
      \begin{center}\begin{large}}{\end{large}\end{center} \end{quotation}}
\def\Acknowledgments{\bigskip  \bigskip \begin{center}
          \large\bf Acknowledgments\end{center}}

\makeatletter
\def\section{\@startsection{section}{0}{\z@}{5.5ex plus .5ex minus
 1.5ex}{2.3ex plus .2ex}{\large\bf}}
\def\subsection{\@startsection{subsection}{1}{\z@}{3.5ex plus .5ex minus
 1.5ex}{1.3ex plus .2ex}{\normalsize\bf}}
\def\subsubsection{\@startsection{subsubsection}{2}{\z@}{-3.5ex plus
-1ex minus  -.2ex}{2.3ex plus .2ex}{\normalsize\sl}}

\renewcommand{\@makecaption}[2]{%
   \vskip 10pt
   \setbox\@tempboxa\hbox{\small #1: #2}
   \ifdim \wd\@tempboxa >\hsize     % IF longer than one line:
       \small #1: #2\par          %   THEN set as ordinary paragraph.
     \else                        %   ELSE  center.
       \hbox to\hsize{\hfil\box\@tempboxa\hfil}
   \fi}

 \def\citenum#1{{\def\@cite##1##2{##1}\cite{#1}}}
\newcount\@tempcntc
\def\@citex[#1]#2{\if@filesw\immediate\write\@auxout{\string\citation{#2}}\fi
  \@tempcnta\z@\@tempcntb\m@ne\def\@citea{}\@cite{\@for\@citeb:=#2\do
    {\@ifundefined
       {b@\@citeb}{\@citeo\@tempcntb\m@ne\@citea\def\@citea{,}{\bf ?}\@warning
       {Citation `\@citeb' on page \thepage \space undefined}}%
    {\setbox\z@\hbox{\global\@tempcntc0\csname b@\@citeb\endcsname\relax}%
     \ifnum\@tempcntc=\z@ \@citeo\@tempcntb\m@ne
       \@citea\def\@citea{,}\hbox{\csname b@\@citeb\endcsname}%
     \else
      \advance\@tempcntb\@ne
      \ifnum\@tempcntb=\@tempcntc
      \else\advance\@tempcntb\m@ne\@citeo
      \@tempcnta\@tempcntc\@tempcntb\@tempcntc\fi\fi}}\@citeo}{#1}}
\def\@citeo{\ifnum\@tempcnta>\@tempcntb\else\@citea\def\@citea{,}%
  \ifnum\@tempcnta=\@tempcntb\the\@tempcnta\else
  {\advance\@tempcnta\@ne\ifnum\@tempcnta=\@tempcntb \else\def\@citea{--}\fi
    \advance\@tempcnta\m@ne\the\@tempcnta\@citea\the\@tempcntb}\fi\fi}
\makeatother

\begin{document}
\begin{titlepage}
\pubblock

\vfill
\def\thefootnote{\fnsymbol{footnote}}
\boldmath
\Title{The width difference of $B_s$ mesons}
\unboldmath
\vfill
\Author{Ulrich Nierste}
\Address{\csumb}
\vfill
\begin{Abstract}
  Next-to-leading order QCD corrections to the width difference \dg\ 
  in the $B_s$-meson system are presented.  I further discuss how \dg\ 
  can be used to detect new physics.
\end{Abstract}
\vfill
\begin{Presented}
5th International Symposium on Radiative Corrections \\ 
(RADCOR--2000) \\[4pt]
Carmel CA, USA, 11--15 September, 2000
\end{Presented}
\vfill
\end{titlepage}
\def\thefootnote{\arabic{footnote}}
\setcounter{footnote}{0}

\section{Introduction}\label{sect:intro}
Currently the prime focus of experimental elementary particle physics
is the investigation of the flavor sector of the Standard Model.
Transitions between different fermion generations originate from the
Higgs-Yukawa sector, which is poorly tested so far. The experimental
effort is not only devoted to a precise determination of
the Cabibbo-Kobayashi-Maskawa (CKM) matrix \cite{ckm}, which parameterizes
the flavor-changing couplings.  Flavor-changing neutral currents
(FCNC) also provide an ideal testing ground to search for new physics,
because they are highly suppressed in the Standard Model: FCNC's are
loop-induced, involve the weak coupling constant and the heavy $W$
boson, are suppressed by small CKM elements or the GIM mechanism
\cite{gim} and further often suffer from a helicity-suppression,
because flavor-changing couplings only involve left-handed fields.
Therefore experiments in flavor physics are much more sensitive to new
physics than the precision tests of the gauge sector performed in the
LEP/SLD/Fermilab-Run-I era.  Decays of $B$ mesons are especially
interesting: they allow us to determine three of the four CKM
parameters, their rich decay spectrum helps to overconstrain the CKM
matrix, they have theoretically clean CP asymmetries (as opposed to
$K\to \pi \pi$ decays), information from $B_d$, $B_s$ and $B^+$ decays
can be combined using $SU(3)_F$ symmetry, the large $b$ quark mass
permits the use of heavy quark symmetries and the heavy quark 
expansion, and in many extensions of the Standard Model third
generation fermions are most sensitive to new physics.
 
\begin{figure}[t!]
\begin{center}
\epsfig{file=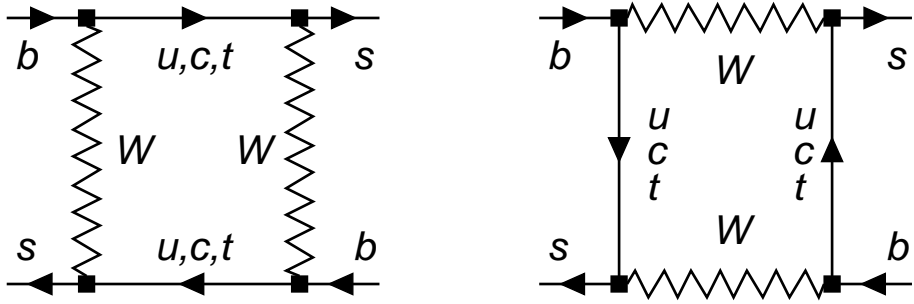,width=0.8\textwidth}
\caption[0]{\label{fig:box} 
Lowest order contribution to \bbms\ in the Standard Model.}
\end{center}
\end{figure}

While $B_s$ mesons cannot be studied at the $B$ factories running on
the $\Upsilon(4S)$ resonance \cite{babar}, they are copiously produced
at hadron colliders \cite{bhad}. $B_s$ mesons mix with their
antiparticles. Therefore the two mass eigenstates $B_H$ and $B_L$ (for
``heavy'' and ``light''), which are linear combinations of $B_s$ and
$\ov{B}_s$, differ in their mass and width.  In the Standard Model
\bbms\ is described in the lowest order by the box diagrams depicted in
Fig.~\ref{fig:box}. 
The dispersive part of the \bbms\ amplitude is called $M_{12}$. In the
Standard Model it is dominated by box diagrams with internal top
quarks. The absorptive part is denoted by $\Gamma_{12}$ and mainly
stems from box diagrams with light charm quarks. $\Gamma_{12}$ is
generated by decays into final states which are common to $B_s$ and
$\ov{B}_s$. While $M_{12}$ can receive sizable contributions from new
physics, $\Gamma_{12}$ is induced by the CKM-favored tree-level decay
$b \to c \ov{c} s$ and is insensitive to new physics.  Experimentally
\bbms\ manifests itself in damped oscillations between the $B_s$ and
$\ov{B}_s$ states.  We denote the mass and width differences between
$B_H$ and $B_L$ by
\beqa%
\dm &=& M_H - M_L \,, \qquad \dg \; = \; \Gamma_L - \Gamma_H . \no
\eeqan%
By solving the eigenvalue problem of $M_{12}-i \Gamma_{12}/2$ one
can relate \dm\ and \dg\ to $M_{12}$ and $\Gamma_{12}$: 
\beqa%
\dm &=&2\, |M_{12}|, \qquad \quad \dg \; =\; 2\, |\Gamma_{12}| \cos
\phi, \label{dgsol} 
\eeqan%
where $\phi$ is defined as 
\beqa%
\frac{M_{12}}{\Gamma_{12}} = - \lt|\frac{M_{12}}{\Gamma_{12}} \rt|\,
e^{i \phi} \label{defphi}.  
\eeqan%
\dm\ equals the \bbs\ oscillation frequency and has not been measured
yet.  In deriving \eq{dgsol} terms of order $|\Gamma_{12}/M_{12}|^2$
have been neglected. $\phi$ in \eq{defphi} is a CP-violating phase,
which is tiny in the Standard Model, so that
$\dg_{\rm SM}=2|\Gamma_{12}|$.  Unlike in the case of $B_d$ mesons, the
Standard Model predicts a sizable width difference \dg\ in the $B_s$
system, roughly between 5 and 30\% of the average total width
$\Gamma=(\Gamma_L+\Gamma_H)/2$.  The decay of an untagged $B_s$
meson into the final state $f$ is in general governed by two
exponentials:
\beqa%
\guntf &\propto& 
e^{-\Gamma_L t} \lt| \langle f \ket{B_L} \rt|^2 + e^{-\Gamma_H t} \lt|
\langle f \ket{B_H} \rt|^2 . \label{twoex} 
\eeqan% 
If $f$ is a flavor-specific final state like $D_s^- \pi^+$ or $X
\ell^+ \nu$, the coefficients of the two exponentials in \eq{twoex}
are equal. A fit of the corresponding decay distribution to a single
exponential then determines the average width $\Gamma$ up to
corrections of order $(\dg)^2/\Gamma$. In the Standard Model CP
violation in \bbms\ is negligible, so that we can simultaneously
choose $B_L$ and $B_H$ to be CP eigenstates and the $b \to c\ov{c} s$
decay to conserve CP.  Then $B_H$ is CP-odd and cannot decay into a
CP-even double-charm final state $f_{CP+}$ like $(J/\psi
\phi)_{L=0,2}$, where $L$ denotes the quantum number of the orbital
angular momentum. Thus a measurement of the $B_s$ width in $B_s \to
f_{CP+}$ determines $\Gamma_L$. By comparing the two measurements one
finds $\dg/2$. CDF will perform this measurement with $B_s \to D_s^-
\pi^+$ and $B_s \to J/\psi \phi $ in Run-II of the Tevatron \cite{mm}.

\section{QCD corrections}
Weak decays of $B$ mesons involve a large range of different mass
scales: first there is the $W$ boson mass $M_W$, which appears in the
weak $b\to c\ov{c} s$ decay amplitude. The second scale in the problem
is the mass $m_b$ of the decaying $b$ quark. Finally there is the QCD
scale parameter $\lqcd$, which sets the scale for the strong binding
forces in the $B_s$ meson. QCD corrections associated with these
scales must be treated in different ways. To this end one employs a
series of operator product expansions, which factorize the studied
amplitude into short-distance Wilson coefficient and matrix elements
of local operators, which comprise the long-distance physics. Here in the
first step the $W$-mediated $b\to c\ov{c} s$ decay amplitude is matched
to matrix elements of local four-quark operators. We need the two 
\dbo\ current-current operators 
\begin{equation}\label{q1q2}
Q_1=  \bar c_i \gamma_\mu (1-\gamma_5) b_j 
      \bar s_j \gamma^\mu (1-\gamma_5) c_i \qquad
Q_2=  \bar c_i  \gamma_\mu (1-\gamma_5) b_i 
      \bar s_j \gamma^\mu (1-\gamma_5) c_j,
\end{equation}
where $i,j$ are color indices. $Q_2$ is pictorially obtained by
contracting the $W$ line in the $b\to c\ov{c} s$ amplitude to a point.
$Q_1$ emerges, once gluon exchange between the two quark lines is
included. In the effective hamiltonian 
\begin{equation}\label{hpeng}
{\cal H}_{eff}=\frac{G_F}{\sqrt{2}} V_{cb}V^*_{cs}
            \sum^2_{r=1} C_r Q_r
\end{equation} 
the Wilson coefficients $C_r$ are determined in such a way that the
Standard Model amplitude is reproduced by $\bra{c \ov{c} s} {\cal
  H}_{eff} \ket{b}$ up to terms of order $m_b^2/M_W^2$. The Fermi
constant $G_F$ and the CKM elements have been factored out in
\eq{hpeng}. The $C_r$'s contain the short-distance physics associated
with the scale $M_W$. QCD corrections to the Wilson coefficients can
be computed in perturbation theory. The renormalization group
evolution of the $C_r$'s down to the scale $\mu_1={\cal O}(m_b)$ sums
the large logarithms $\ln (\mu_1/M_W)$ to all orders in perturbation
theory. The minimal way to do this is the leading log approximation
which reproduces all term of order $\alpha_s^n \ln^n (\mu_1/M_W)$,
$n=0,1,\ldots$, of the full Standard Model transition amplitude.  The
next-to-leading order (NLO) corrections to the coefficients comprise
the terms of order $\alpha_s^{n+1} \ln^n (\mu_1/M_W)$ and have been
calculated in \cite{bw}. We remark that there are also penguin
operators in the effective hamiltonian ${\cal H}_{eff}$. We have
omitted them in \eq{hpeng}, because their coefficients are very
small. Their impact is discussed in \cite{bbd,bbgln}. 

\begin{figure}[t]
\centerline{
\epsfxsize=0.8\textwidth 
\epsfbox{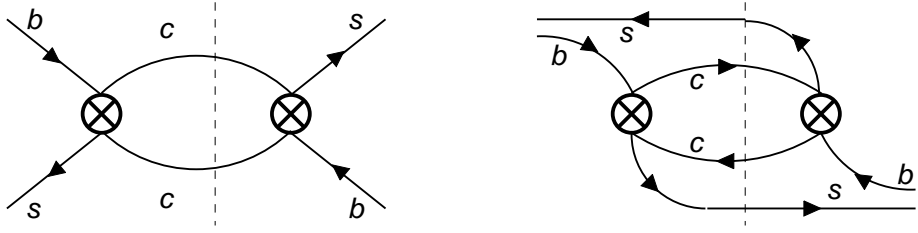}}
\caption{Leading-order diagrams for $\Gamma_{12}$}\label{fig:b2}
\end{figure}

$\dg_{\rm SM}=2|\Gamma_{12}|$ is related to ${\cal H}_{eff}$ by the
optical theorem:
\begin{eqnarray}
\dg_{\rm SM} & = & 2 |\Gamma_{12}|  
\; =  \; \lt| - \frac{1}{M_{B_s}}\,
\mbox{Abs}\, \langle\bar B_s| 
\,i\int d^4x\ T\,{\cal H}_{eff}(x){\cal H}_{eff}(0)
|B_s\rangle \rt| . \label{dgopt}
\end{eqnarray} 
Here `Abs' denotes the absorptive part of the amplitude, which is
obtained by retaining only the imaginary part of the loop integration.
The corresponding leading-order diagrams are shown in
Fig.~\ref{fig:b2}.
In the next step of our calculation we perform an operator product 
expansion of the RHS in \eq{dgopt} in order to describe $\Gamma_{12}$
in terms of matrix elements of local \dbt\ operators:
\begin{eqnarray}
\lefteqn{
\hspace{-2ex}
| \mbox{Abs}\, \langle\bar B_s| 
\,i\int d^4x\ T\,{\cal H}_{eff}(x){\cal H}_{eff}(0)
|B_s\rangle | } \nn
&=& -\frac{G^2_F m^2_b}{12\pi} \lt| V^*_{cb}V_{cs} \rt|^2
\, \cdot \nn
&&\!\!
\left[ F\lt( \frac{m_c^2}{m_b^2} \rt) 
\langle\bar B_s| Q |B_s\rangle + 
F_S \lt( \frac{m_c^2}{m_b^2} \rt)  \langle\bar B_s| Q_S |B_s\rangle \right] 
\lt[1+ {\cal O} \lt( \frac{\lqcd}{m_b} \rt) \rt] \!\! .
\label{dghqe}
\end{eqnarray}
The two dimension-6 operators appearing in 
\eq{dghqe} are 
\begin{equation}\label{qqs}
Q = \bar s_i \gamma_\mu (1-\gamma_5) b_i 
    \bar s_j \gamma^\mu (1-\gamma_5)b_j ,\qquad
Q_S= \bar s_i (1+\gamma_5) b_i  \bar s_j (1+\gamma_5) b_j  .
\end{equation}
In the leading order of QCD the RHS of \eq{dghqe} is pictorially
obtained by simply shrinking the $(c,\ov{c})$ loop in
Fig.~\ref{fig:b2} to a point. Our second operator product expansion is
also called \emph{heavy quark expansion}\ (HQE), which has been
developed long ago by Shifman and Voloshin \cite{hqe}.  The new Wilson
coefficients $F$ and $F_S$ also depend on the charm quark mass $m_c$,
which is formally treated as a hard scale of order $m_b$, since $m_c
\gg \lqcd$. Strictly speaking, the HQE in \eq{dghqe} is an expansion
in $\lqcd/\sqrt{m_b^2-4 m_c^2}$.  For the calculation of $F$ and $F_S$
it is crucial that these coefficients do not depend on the infrared
structure of the process. In particular they are independent of the
QCD binding forces in the external $B_s$ and $\ov{B}_s$ states in
\eq{dghqe}, so that they can be calculated in perturbation theory at
the parton level. The non-perturbative long-distance QCD effects
completely reside in the hadronic matrix elements of $Q$ and $Q_S$.
It is customary to parametrize these matrix as
\begin{eqnarray}
\langle\bar B_s|Q(\mu_2)|B_s\rangle &=& 
 \frac{8}{3}f^2_{B_s}M^2_{B_s} B (\mu_2)
\nn
\langle\bar B_s|Q_S(\mu_2)|B_s\rangle &=& -\frac{5}{3}f^2_{B_s}M^2_{B_s}
\frac{M^2_{B_s}}{(m_b(\mu_2)+m_s(\mu_2))^2}  B_S (\mu_2) . \label{me}
\end{eqnarray}
Here $M_{B_s}$ and $f_{B_s}$ are mass and decay constant of the $B_s$
meson. The quark masses $m_b$ and $m_s$ in \eq{me} are defined in the
$\ov{\rm{MS}}$ scheme.  In the so called vacuum insertion
approximation $B(\mu_2)$ and $B_S(\mu_2)$ are equal to 1.
$\mu_2={\cal O}(m_b)$ is the scale at which the \dbt\ operators are
renormalized. It can be chosen different from $\mu_1$. The dependence
of \dg\ on the unphysical scales $\mu_1$ and $\mu_2$ diminishes
order-by-order in perturbation theory. The residual dependence is
usually used as an estimate of the theoretical uncertainty. The
$\mu_1$-dependence cancels between the \dbo\ Wilson coefficients
$C_{1,2}$ in \eq{hpeng} and the radiative corrections to $F$ and $F_S$
in \eq{dghqe}. The terms in $F$ and $F_S$ which depend on $\mu_2$
cancel with corresponding terms in $B(\mu_2)$ and $B_S(\mu_2)$.  The
scale $\mu_2$ enters a lattice calculation of these non-perturbative
parameters when the lattice quantities are matched to the continuum.

\begin{figure}[t]
\centerline{
\epsfysize=0.95\textwidth 
\rotate[r]{\epsfbox{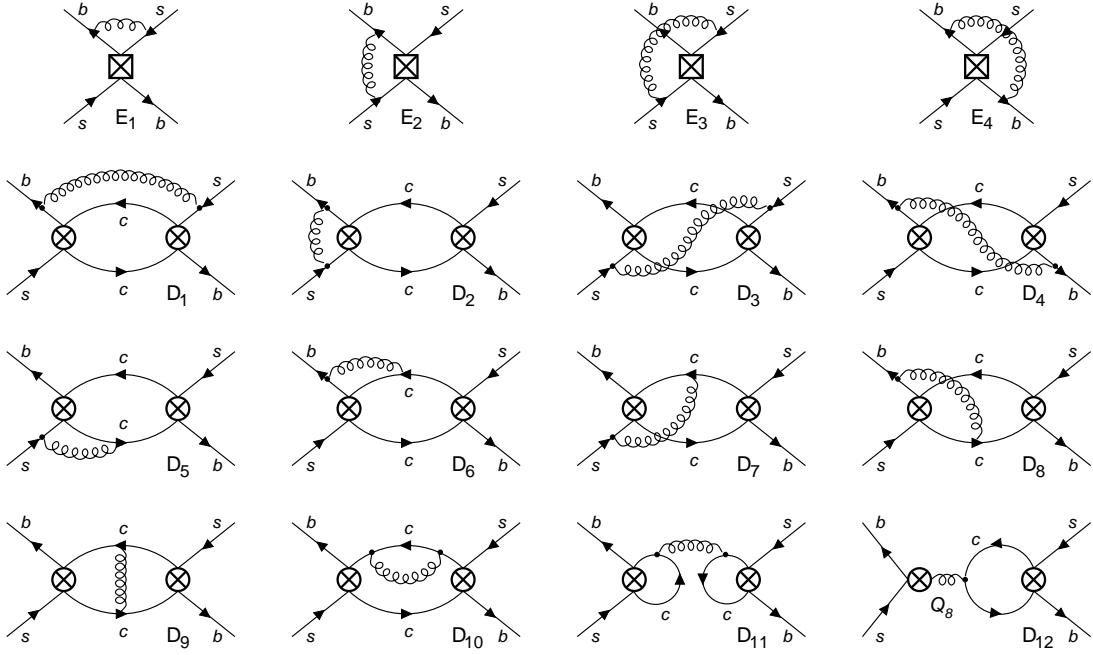}}}
\caption{Next-to-leading-order diagrams for $\Gamma_{12}$.
$Q_8$ is the chromomagnetic penguin operator.}\label{fig:b3}
\end{figure}

The leading-order calculation of \dg\ requires the calculation of the
diagrams in Fig.~\ref{fig:b2} and has been performed long ago
\cite{LO}. Subsequently corrections of order $\lqcd/m_b$ to \eq{dghqe}
have been computed in \cite{bbd}.  The next-to-leading order
calculation requires the calculation of the diagrams depicted in
Fig.~\ref{fig:b3} \cite{bbgln}.
The motivations for this cumbersome calculation are
\begin{itemize}
\item[1)] to verify the infrared safety of $F$ and $F_S$,
\item[2)] to allow for an experimental test of the HQE,
\item[3)] a meaningful use of lattice results for 
        hadronic matrix elements,
\item[4)] a consistent use of $\Lambda_{\ov{\textit{\scriptsize{MS}}}}$,
\item[5)] to reduce the sizable $\mu_1$-dependence of the LO,
\item[6)] the large size of QCD corrections, typically of order $30\% $. 
\end{itemize}
The disappearence of infrared effects from the Wilson coefficients $F$
and $F_S$ mentioned in point 1)\ is necessary for any meaningful
operator product expansion. Yet early critics of the HQE had found
power-like infrared divergences in individual cuts of diagrams of
Fig.~\ref{fig:b3}. In response the cancellation of these divergences
has been shown \cite{bu}, long ago before we have performed the full
NLO calculation. However, there are also logarithmic infrared
divergences. We found IR-singularities to cancel via two mechanisms:
\begin{itemize}
\item Bloch-Nordsiek cancellations among different cuts of the 
        same diagram,
\item factorization of IR-singularities, which end up in 
         $ \langle \bar B_s | Q   | B_s \rangle$, 
   $ \langle \bar B_s | Q_S | B_s \rangle$.
\end{itemize}
Point 2) above addresses the conceptual basis of the HQE, which is
sometimes termed \emph{quark-hadron duality}. It is not clear, whether
the HQE reproduces all QCD effects completely.  Exponential terms like
$\exp (-\kappa m_b/\lqcd)$, for example, cannot be reproduced by a
power series \cite{s}. The relevance of such terms can at present only
be addresses experimentally, by confronting HQE-based predictions with
data. The only QCD information contained in the LO prediction for
\dg\ is the coefficients of $\alpha_s^n \ln^n M_W$, associated with
hard gluon exchange along the $W$-mediated $b \to c\ov{c} s$
amplitude. The question of quark-hadron duality, however, addresses
the non-logarithmic QCD corrections, which belong to the NLO. While it
is certainly very interesting to find violations of quark-hadron
duality in $B$ physics, it will be hard to detect them in \dg: in the
LO diagrams in Fig.~\ref{fig:b2} the heavy $c$,$\ov{c}$ quarks recoil
back-to-back against each other and are fast in the $b$ rest frame.
The inclusive $b\to c \ov{c} s$ decay is more sensitive to
uncontrolled long-distance effects, because in some parts of the phase
space the $c$,$\ov{c}$ quarks move slowly in the $b$ rest frame or
with respect to each other. Still the HQE prediction for $b\to c
\ov{c} s$ \cite{bbfg} agrees with experiment \cite{bccsexp}. If one
further takes into account that \dg\ has an overall hadronic
uncertainty associated with $f_{B_s}^2 B$ and $f_{B_s}^2 B_S$, it
appears very unlikely that violations of quark-hadron duality can be
detected in \dg. Points 3) and 4) are related to the fact that
leading-order predictions are not sensitive to the renormalization
scheme, which impedes the lattice-continuum matching of the
non-perturbative parameters. Likewise the $\mu_2$ dependence of this
matching procedure cannot be addressed in the leading-order. The
$\mu_{1}$-dependence of \dg\ is huge in the leading order. It is
reduced in the NLO, but still remains sizable. The results for $F$ and
$F_S$ can be found in Tab.~\ref{tab}.  The reduction of the $\mu_1$
dependence can be verified from the table. The numerical values of the
NLO coefficients depend on the renormalization scheme. The precise
definition of this scheme involves the subtraction prescription for
the ultraviolet poles (dimensional regularization with $\ov{\rm{MS}}$
subtraction \cite{bbdm}), the treatment of $\gamma_5$ (for which we
have used the NDR scheme) and the chosen definitions of the evanescent
operators \cite{hn}, which can be found in \cite{bbgln}. The
lattice-continuum matching must be done in the same renormalization
scheme, so that all scheme dependences cancel in the prediction for
\dg.

\def\tableline{\noalign{%\vskip-.5pt
\hrule height.7pt depth0pt\vskip3pt}}
\begin{table}[tb]
\caption{\label{tab:wc}
}
\begin{center}
\setlength{\tabcolsep}{9pt}
\renewcommand{\arraystretch}{1.2}
\begin{displaymath}
\begin{array}{cccc} 
\tableline
\mu_1 & m_b/2 & m_b & 2 m_b \\ \hline\hline
-F_S                  & 0.867 & 1.045 & 1.111  \\  \hline
-F^{(0)}_S            & 1.729 & 1.513 & 1.341 \\  \hline
F                    & 0.042 & 0.045 & 0.049  \\  \hline
F^{(0)}              & 0.030 & 0.057 & 0.103 \\  \hline
\end{array}
\end{displaymath}
\caption{\label{tab}
  Numerical values of the Wilson coefficients $F$ and $F_S$ for
  $m_c^2/m_b^2=0.085$. Leading-order results are indicated with the
  superscript $(0)$.  The precise definition of our renormalization
  scheme can be found in \cite{bbgln}.}
\end{center}
\end{table}
%%%%%%%%%%%%%%%%%%%%%%%%%%%%%%%%%%%%%%%%%%%%%%%%%%%%%%%%%%%%%%%%%%%%%%%%%

Including the corrections of order $\lqcd/m_b$ \cite{bbd} our
NLO prediction reads
\begin{eqnarray} 
\frac{\dg_{\rm SM}}{\Gamma} &=& 
        \left( \frac{f_{B_s}}{245~{\rm MeV}} \right)^2 \,
        \left[ \, (0.234\pm 0.035)\, B_S (m_b) - 0.080 \pm 0.020 \, \right] .
\label{dgnum} 
\end{eqnarray} 
Here $m_b(m_b)+m_s(m_b)=4.3\,$GeV (in the $\ov{\rm MS}$ scheme) and
$m_c^2/m_b^2=0.085$ has been used. Since $F$ is small, the uncertainty
in $B$ is irrelevant, and the term involving $FB$ has been absorbed
into the constant $ - 0.080 \pm 0.020$ in \eq{dgnum}.  Recently the
KEK--Hiroshima group succeeded in calculating $f_{B_s}$ in an
unquenched lattice QCD calculation with two dynamical fermions
\cite{fbs}. The result is $f_{B_s}=(245\pm 30)\,$MeV.  A recent
quenched lattice calculation has found $B_S(m_b)=0.87 \pm 0.09$
\cite{hioy} for the $\ov{\rm MS}$ scheme. A similar result has been
obtained in \cite{bmrglm}. 
With these numbers one finds from \eq{dgnum}:
\begin{eqnarray} 
\frac{\dg_{\rm SM}}{\Gamma} &=& 0.12\pm 0.06 
        . \label{dgnum2}
\end{eqnarray}
Here we have conservatively added the errors from the two lattice
quantities linearly.

\section{New physics}
% The two extreme new physics scenarios studied in $B$ physics are 
% \begin{itemize} 
% \item[1)] models, in which the flavor structure stems solely from 
%           Standard-Model like Yukawa interactions\\
%           and 
% \item[2)] models with new sources of FCNCs and in particular new 
%           sources of CP violation.    
% \end{itemize}
% The Constrained Minimal Supersymmetric Standard Model, for example,
% belongs to the first class, while generic supersymmetric models 
% (with flavor-nondiagonal quark-squark-gluino couplings) are in class
% 2). In both cases \dg\ can provide useful information. 

% In the near future the unitarity triangle will be precisely
% constructed from the CP asymmetry in $B_d \to \psi K_S$, which
% determines $\sin(2 \beta)$, and the ratio $\dm_d/\dm_s$.  In class 1)
% models these two quantities are unaffected by new physics and
% therefore they determine the true unitarity triangle. The new physics
% also drops out from other CP-asymmetries and from $|V_{ub}/V_{cb}|$
% measured in semileptonic $b$ decays. Observables which are sensitive
% to new physics are therefore 

% The phase CP-violating phase $\phi$ 
% in \eq{defphi}  
In the presence of new physics $\arg M_{12}$ and thereby $\phi$ in
\eq{defphi} can assume any value. Non-standard contributions to $\phi$
can be measured from CP-asymmetries, which requires the resolution of
the rapid \bbs\ oscillations and tagging, i.e.\ the discrimination
between $B_s$ and $\ov{B}_s$ mesons at the time $t=0$ of their
production. From \eq{dgsol} one verifies that a non-vanishing $\phi$
also affects \dg, which can be measured from untagged data samples and
therefore involves better efficiencies than tagged studies.  Of course
in the search for new physics \dg\ is only competitive with CP
asymmetries, which determine $\sin \phi$, if $\phi$ is not too close
to 0 or $\pm\pi$. Nevertheless the information on $\phi$ from both tagged
and untagged data should be combined.

As discussed at the end of Sect.~\ref{sect:intro}, \dg\ is most easily
found from the lifetimes measured in the decays of an untagged $B_s$
sample into a flavor-specific final state and into a CP-specific final
state $f_{CP}$, respectively.  In the presence of a non-zero
CP-violating phase $\phi$ the mass eigenstates $B_L$ and $B_H$ are no
more CP eigenstates, so that now both exponentials in \eq{twoex}
contribute to the decay $B_s \to f_{CP}$.  Then this method determines
\cite{g,dfn}:
\beqa%
\dg \cos\phi &=& \dg_{\rm SM} \cos^2 \phi . \label{dgnp} 
\eeqan% 
As first pointed out in \cite{g}, one can determine $|\cos \phi|$
without using the theoretical input in \eq{dgnum}: if one is able to
resolve both exponentials of \eq{twoex} in the time evolution of a
$B_s$ decay into a flavor-specific final state, one will measure the true 
$|\dg|$. By comparing with \eq{dgnp} one can then solve for $|\cos\phi|$. 
This method, however, requires to distinguish $\cosh ((\dg )t/2)$ from
1 and is very difficult to carry out. In \cite{dfn} a different method
has been proposed, which only requires to measure lifetimes and
branching ratios: first define CP eigenstates
$B_s^{\textrm{\scriptsize odd}}$ and $B_s^{\textrm{\scriptsize even}}$ 
such that $B_s^{\textrm{\scriptsize odd}}\to 
\hspace{-2.1ex}{\scriptstyle /} \hspace{1.5ex} D_s^+ D_s^-$. Then define
\begin{eqnarray}
\dg_{\rm CP} & = & 
\Gamma \lt( B_s^{\textrm{\scriptsize even}} \rt) - 
        \Gamma ( B_s^{\textrm{\scriptsize odd}} ) .
        \label{dgcp2}
\end{eqnarray}
$\dg_{\rm CP}$ is related to $\Gamma_{12}$ as
\begin{eqnarray}
\dg_{\rm CP} & = &  2 |\Gamma_{12}| . \no
\end{eqnarray}
Hence $\dg_{\rm CP}$ equals $\dg_{\rm SM}$, but is not affected by the new
physics phase $\phi$ at all! By measuring both $\dg_{\rm CP}$ and $\dg
\cos\phi$ one can infer $|\cos \phi|$ from \eq{dgnp}. Loosely
speaking, $\dg_{\rm CP}$ is measured by counting the CP-even and
CP-odd double-charm final states in $B_s$ decays:
\begin{eqnarray} 
\dg_{\rm CP} & = & 
 2\,  \Gamma \sum_{f \in X_{c\ov{c}}}  \brunts{f} 
        \, 
        (1- 2\, x_f)\, \lt[ 1 \, + \, {\cal O} \lt( \frac{\dg}{\Gamma}
        \rt) \rt].
\label{dgcpres} 
\end{eqnarray}
Here \brunts{f}\ is the branching ratio of an untagged $B_s$ meson
into the final state $f$, $\Gamma$ is the average $B_s$ width, the sum
runs over all double-charm final states and $x_f$ is the CP-odd
component of the final state $f$, e.g.\ $x_f$ is 0 for a CP-even state
and equals 1 for a CP-odd state.  In the Shifman-Voloshin limit
\cite{sv} one can show that $\dg_{\rm CP}$ is exhausted by the
$D_s^{(*)}{}^+ D_s^{(*)}{}^- $ final states \cite{ayopr}.  Moreover
these four final states are purely CP-even in this limit. ALEPH has
measured the sum of these branching ratios \cite{aleph} and found,
relying on the SV limit,
\begin{eqnarray}
\dg_{\rm CP} \; \approx \; 
 2\, \brunts{D_s^{(*)}{}^+ D_s^{(*)}{}^-} & = & 
        0.26 \epm{0.30}{0.15} \label{alexp} .
\end{eqnarray} 
In the future one can extend this method by including all detected
double-charm final states into the sum in \eq{dgcpres} and determine
the CP-odd fraction $x_f$ of each final state by measuring the $B_s$
lifetime in the studied mode \cite{dfn}.

\Acknowledgments
I thank Gudrun Hiller for inviting me to this conference. I gratefully
appreciate the financial support from the conference.

\end{document}